\documentclass[12pt,preprint]{aastex}
\usepackage{ulem}
\usepackage{color}
\usepackage{amsmath}

\begin{document}
\title{KamLAND Sensitivity to Neutrinos from Pre-Supernova Stars}

\author{
K.~Asakura\altaffilmark{1}, 
A.~Gando\altaffilmark{1}, 
Y.~Gando\altaffilmark{1}, 
T.~Hachiya\altaffilmark{1},  
S.~Hayashida\altaffilmark{1}, 
H.~Ikeda\altaffilmark{1},
K.~Inoue\altaffilmark{1,2},
K.~Ishidoshiro\altaffilmark{1},
T.~Ishikawa\altaffilmark{1}
S.~Ishio\altaffilmark{1}
M.~Koga\altaffilmark{1,2},
S.~Matsuda\altaffilmark{1},
T.~Mitsui\altaffilmark{1},
D.~Motoki\altaffilmark{1},
K.~Nakamura\altaffilmark{1,2}, 
S.~Obara\altaffilmark{1}, 
T.~Oura\altaffilmark{1}, 
I.~Shimizu\altaffilmark{1}, 
Y.~Shirahata\altaffilmark{1}, 
J.~Shirai\altaffilmark{1},
A.~Suzuki\altaffilmark{1}, 
H.~Tachibana\altaffilmark{1},
K.~Tamae\altaffilmark{1},
K.~Ueshima\altaffilmark{1},
H.~Watanabe\altaffilmark{1},
B.D.~Xu\altaffilmark{1,18},
A.~Kozlov\altaffilmark{2},
Y.~Takemoto\altaffilmark{2},
S.~Yoshida\altaffilmark{3},
K.~Fushimi\altaffilmark{4},
A.~Piepke\altaffilmark{5,2},
T.~I.~Banks\altaffilmark{6,7},
B.~E.~Berger\altaffilmark{7,2},
B.K.~Fujikawa\altaffilmark{7,2},
T.~O'Donnell\altaffilmark{6,7},
J.G.~Learned\altaffilmark{8},
J.~Maricic\altaffilmark{8}, 
S.~Matsuno\altaffilmark{8}, 
M.~Sakai\altaffilmark{8},
L.~A.~Winslow\altaffilmark{9},
Y.~Efremenko\altaffilmark{10,11,2},
H.~J.~Karwowski\altaffilmark{12,13},
D.~M.~Markoff\altaffilmark{12,14},
W.~Tornow\altaffilmark{12,15,2},
J.~A.~Detwiler\altaffilmark{16},
S.~Enomoto\altaffilmark{16,2},
M.P.~Decowski\altaffilmark{17,2}
}

\affil{The KamLAND Collaboration}

\altaffiltext{1}{Research Center for Neutrino Science, Tohoku University, Sendai 980-8578, Japan}
\altaffiltext{2}{Kavli Institute for the Physics and Mathematics of the Universe (WPI), The University of Tokyo Institutes for Advanced Study, The University of Tokyo, Kashiwa, Chiba 277-8583, Japan}
\altaffiltext{3}{Graduate School of Science, Osaka University, Toyonaka, Osaka 560-0043, Japan}
\altaffiltext{4}{Faculty of Integrated Arts and Science, University of Tokushima, Tokushima, 770-8502, Japan}
\altaffiltext{5}{Department of Physics and Astronomy, University of Alabama, Tuscaloosa, Alabama 35487, USA}
\altaffiltext{6}{Physics Department, University of California, Berkeley, California 94720, USA} 
\altaffiltext{7}{Lawrence Berkeley National Laboratory, Berkeley, California 94720, USA}
\altaffiltext{8}{Department of Physics and Astronomy, University of Hawaii at Manoa, Honolulu, Hawaii 96822, USA}
\altaffiltext{9}{Massachusetts Institute of Technology, Cambridge, Massachusetts 02139, USA}
\altaffiltext{10}{Department of Physics and Astronomy, University of Tennessee, Knoxville, Tennessee 37996, USA}
\altaffiltext{11}{National Research Nuclear University, Moscow, Russia}
\altaffiltext{12}{Triangle Universities Nuclear Laboratory, Durham, North Carolina 27708, USA}
\altaffiltext{13}{The University of North Carolina at Chapel Hill, Chapel Hill, North Carolina 27599, USA}
\altaffiltext{14}{North Carolina Central University, Durham, North Carolina 27701, USA}
\altaffiltext{15}{Physics Departments at Duke University, Durham, North Carolina 27705, USA}
\altaffiltext{16}{Center for Experimental Nuclear Physics and Astrophysics, University of Washington, Seattle, Washington 98195, USA}
\altaffiltext{17}{Nikhef and the University of Amsterdam, Science Park, Amsterdam, the Netherlands}
\altaffiltext{18}{Current address: Kavli Institute for the Physics and Mathematics of the Universe (WPI), University of Tokyo, Kashiwa 277-8568, Japan}

\vspace{+0.2in}

\begin{abstract}
In the late stages of nuclear burning for massive stars ($M>8~M_{\sun}$), the production of neutrino-antineutrino pairs through various processes becomes the dominant stellar cooling mechanism. As the star evolves, the energy of these neutrinos increases and in the days preceding the supernova a significant fraction of emitted electron anti-neutrinos exceeds the energy threshold for inverse beta decay on free hydrogen. This is the golden channel for liquid scintillator detectors because the coincidence signature allows for significant reductions in background signals. We find that the kiloton-scale liquid scintillator detector KamLAND can detect these pre-supernova neutrinos from a star with a mass of $25~M_{\sun}$ at a distance less than 690~pc with 3$\sigma$ significance before the supernova. This limit is dependent on the neutrino mass ordering and background levels. KamLAND takes data continuously and can provide a supernova alert to the community. 
\end{abstract}

\keywords{neutrinos, supernovae: general}

\maketitle

\section{Introduction}
The first extrasolar neutrinos were detected from SN1987A by the Kamiokande-II~\citep{hirata1987,hirata1988}, IMB~\citep{IMB}, and Baksan~\citep{baksan} experiments. This dataset has provided many insights into the properties of neutrinos and the physics of supernovae~\citep{Vissani2015}. SN1987A was located in the Large Magellanic Cloud at a distance of $\sim 50$~kpc. A core-collapse supernova in the Milkyway proper would provide a larger flux of neutrinos. This combined with the large suite of running neutrino experiments makes the next Galactic supernova a greatly anticipated event~\citep{kate2012}. 

In a Type II supernova, a huge burst of neutrinos is released, carrying away $\sim10^{53}$~ergs of energy in 10~s. Leading up to this cataclysmic event, neutrinos have already been playing an important role in the cooling of the evolving giant star. Starting in the carbon burning phase, the dominant mechanism for cooling these massive $(M > 8~M_{\sun})$ stars is the loss of energy due to $\nu/\bar{\nu}$ pairs created by thermal processes. 
From applying the discussion in~\cite{Itoh1996} to \cite{Woosley2015}, the dominant process in most of 
$M > 10~M_{\sun}$ stars is the pair process, $e^+e^-\rightarrow \nu \bar{\nu}$.  For other stars with smaller masses, the plasmon decay becomes more important, $\gamma\rightarrow \nu \bar{\nu}$. Secondary contributions come from the photo process, $\gamma e^{-} \rightarrow e^{-}  \nu \bar{\nu}$ and bremsstrahlung, $e^{-}(Ze) \rightarrow (Ze) e^{-}  \nu \bar{\nu}$. 
These thermal processes are often used to set limits on non-standard neutrino interactions since such processes would change the evolution of these objects~\citep{Heger2009}. The most stringent limits on the neutrino magnetic moment come from this type of analysis~\citep{Arceo20151}.

\begin{figure}
\begin{center}
\includegraphics[width=1.0\linewidth]{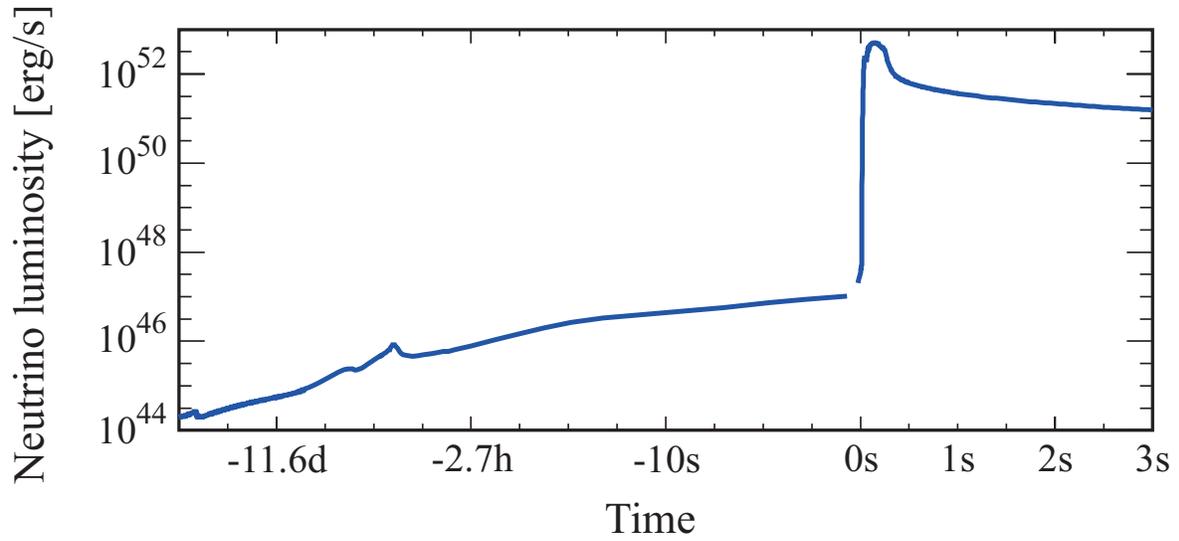}
\end{center}
\caption{Time evolution of the $\bar{\nu}_e$ luminosity of preSN just before collapse~\citep{Odrzywolek2010} 
and of SN after collapse~\citep{nakazato}. Note, the time scale of the horizontal axis which is linear after the collapse but logarithmic before collapse. }
\label{fig1}
\end{figure}

Since these thermal neutrinos precede the supernova, they can also be called pre-supernova neutrinos (preSN). 
Figure~\ref{fig1} shows the overall time evolution of the $\bar{\nu}_e$ luminosity before and after the collapse according to the preSN model developed by the Odrzywolek group~\citep{Odrzywolek2004,Odrzywolek2010}. 
The supernova neutrinos (SN) which follow the collapse based on \cite{nakazato} are also shown for reference. 
Although the preSN luminosity is several orders of magnitude 
smaller than the SN luminosity, the detection of preSN is desired since 
the preSN encode information about the late stages of stellar evolution for high mass stars and could act as a supernova alert, 
a more detailed discussion is found in section \ref{discussion}. 

SN extend to a few tens of MeV. In comparison, the average energy of preSN is low, typically $E < 2$~MeV. 
In this energy range, there are three reactions that can be used to detect these neutrinos in realtime: coherent neutrino scattering, neutrino-electron scattering and inverse beta decay~(IBD), $\bar{\nu}_e + p \rightarrow e^{+} + n$. IBD has one of the highest cross sections for neutrino detection. It also has relatively low backgrounds due to the easily identifiable delayed coincidence signal created by the prompt positron annihilation followed by the delayed neutron capture.  Depending on the detector material, coherent neutrino scattering may have a higher cross section than IBD, but the signal has never been observed due to the very low reconstructed energy of the recoiling nucleus. The detection of preSN through neutrino-electron scattering is possible. However, its cross section is lower than IBD, which reduces the total number of detected events, and the background rate is high since there is no coincidence signal. 
Thus, IBD is the most promising channel for preSN detection. 
 
The energy threshold for IBD is 1.8~MeV. A few days before the supernova, a significant fraction of $\bar{\nu}_e$ exceeds the IBD threshold and it becomes possible to detect the preSN with IBD. IBD is the main supernova channel for both liquid-scintillator detectors and water-Cherenkov detectors like Super-Kamiokande. Water-Cherenkov detectors have relatively high energy thresholds, such as $E_{e}=4.5$~MeV~\citep{Renshaw2014}. This limits both the number of IBD prompt events and the efficiency for detecting the delayed neutron capture.
In comparison, monolithic liquid-scintillator detectors have energy thresholds below 1~MeV and are therefore able to sample a larger fraction of the preSN prompt energy spectrum and effectively detect the neutron capture. Thus, liquid-scintillator detectors have an advantage for the detection of preSN, even if they are smaller than typical water-Cherenkov detectors. 

There are two operating monolithic liquid-scintillator detectors with low-energy thresholds, KamLAND and Borexino~\citep{Cadonati2002361}. 
The SNO+ detector~\citep{sno+} is expected to come online soon and construction has started on the 20~kton JUNO detector~\citep{juno}. In addition, there are several proposals for multi-kton experiments such as RENO-50~\citep{Kim:2014rfa}, HANOHANO~\citep{hanohano}, LENA~\citep{lena2012}, and ASDC~\citep{asdc2014}. All of these detectors would be sensitive to this preSN IBD signal.  A large Gd-doped water-Cherenkov detector such as Gd-doped Super-Kamiokande~\citep{GADZOOKS} would have increased sensitivity due to the higher neutron capture detection efficiency but the higher energy threshold continues to limit the sensitivity. The Baksan and LVD scintillator detectors are similarly limited in their sensitivity to preSN due to their relatively high energy thresholds~\citep{baksan2011,LVD}. 

In the previous studies~\citep{Odrzywolek2004,Odrzywolek2010,kato2015}, the expected number of IBD events in several detectors is evaluated without a detailed detector response model.
We focus on KamLAND since it is currently the largest monolithic liquid-scintillator detector. 
In this article, we quantify KamLAND's sensitivity to the preSN using the actual background rates and a realistic detector response model. 
We discuss the development of a supernova alert based on preSN.  Betelgeuse is a well-known possible supernova progenitor~\citep{Dolan2014} and we evaluate the performance of the preSN alert based on this astrophysical object.

\section{PreSN signal}
\begin{figure}
\begin{center}
\includegraphics[width=1.0\linewidth]{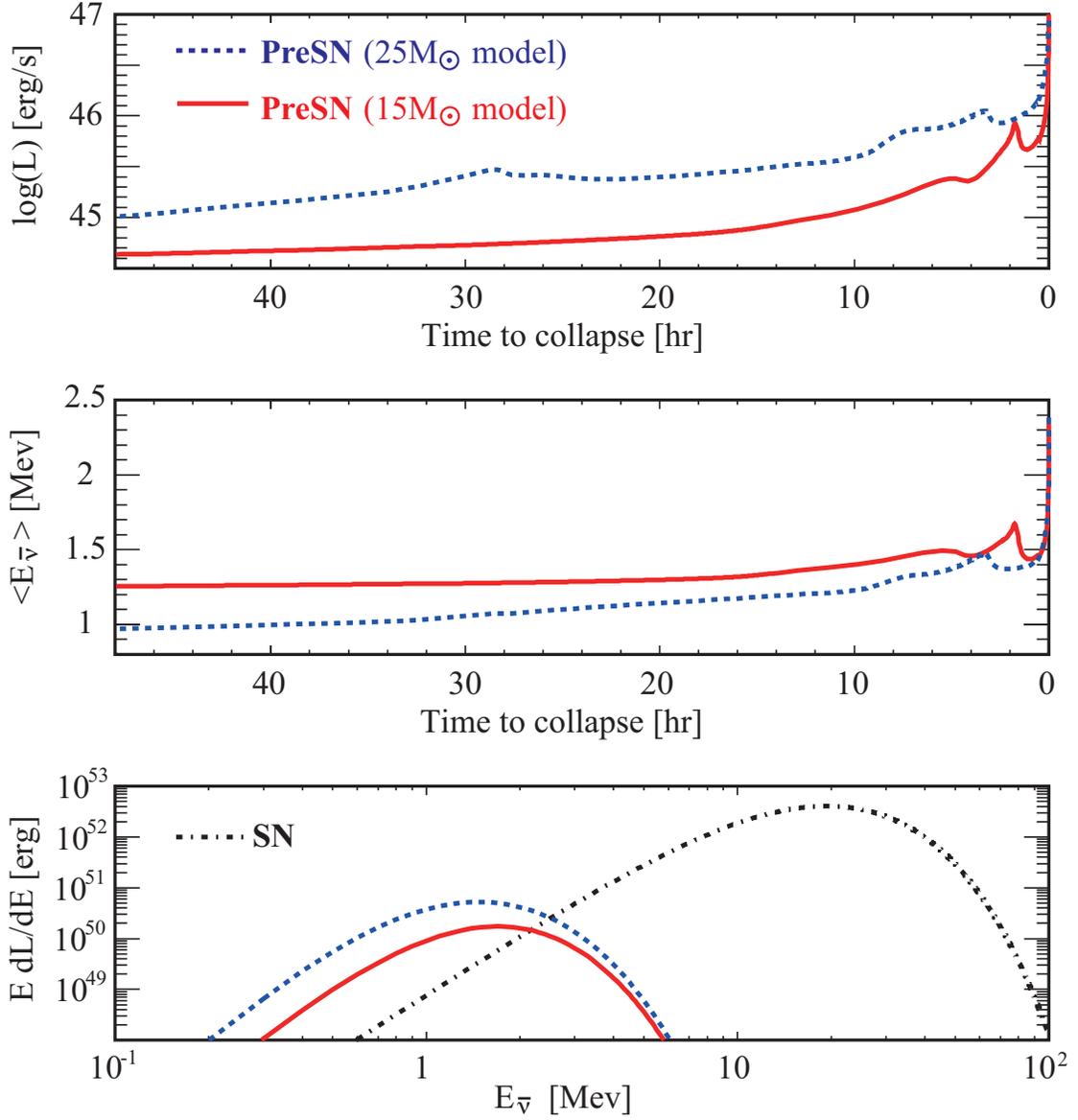}
\end{center}
\caption{
Details of preSN with 15~$M_{\rm \sun}$ and 25~$M_{\rm \sun}$ star models. 
(Top) time evolution of $\bar{\nu}_e$ luminosity, 
(Middle) time evolution of the averaged $\bar{\nu}_e$ energy, 
and (Bottom) differential luminosity weighted by energy ($E_{\bar{\nu}_e} dL/dE_{\bar{\nu}_e} \sim dL/d\log E_{\bar{\nu}_e}$) 
 integrating over 48~hr  preceding collapse with SN integrating over 10~s for reference. SN is calculated from 
a Fermi-Dirac distribution of the flux with a total luminosity of $5\times 10^{52}$\,erg and an average energy of 15~MeV. 
}
\label{fig2}
\end{figure}

The first calculation of the number of detected preSN is found in~\cite{Odrzywolek2004} and updates can be found in~\cite{Odrzywolek2010} and 
\cite{kato2015}.  We use the preSN spectra $\phi_{M}(t,E_{\bar{\nu}_e};d)$ as a function of time and energy from the Odrzywolek's results corrected for the distance of $d$ to the pre-supernova star. We use this to calculate KamLAND's sensitivity to preSN with two example stars of $M=15\,M_{\sun}$ and $M=25\,M_{\sun}$. 
Figure~\ref{fig2} shows the time evolution of the $\bar{\nu}_e$ luminosity in the top panel 
and the averaged $\bar{\nu}_e$ energy in the middle panel during the 48~hr before the collapse. 
The integrated $\bar{\nu}_e$ luminosity over the last 48~hr preceding collapse is $1.9\times 10^{50}$~erg and $6.1\times 10^{50}$~erg, respectively, for the two star masses. 
They correspond to $1.2 \times 10^{56}$ $\bar{\nu}_e$ and $3.8 \times 10^{56}$ $\bar{\nu}_e$, respectively. 
The weighed differential luminosity by energy, $E_{\bar{\nu}_e}dL/dE_{\bar{\nu}_e} \sim dL/d\log E_{\bar{\nu}_e}$, 
is also shown in the bottom of Figure~\ref{fig2} with the SN for reference.  The average energy of the integrated $\bar{\nu}_e$ flux is 1.4~MeV and 1.2~MeV for the $15~M_\sun$ and $25~M_\sun$ models, respectively. 

\begin{figure}
\begin{center}
\includegraphics[width=1.0\linewidth]{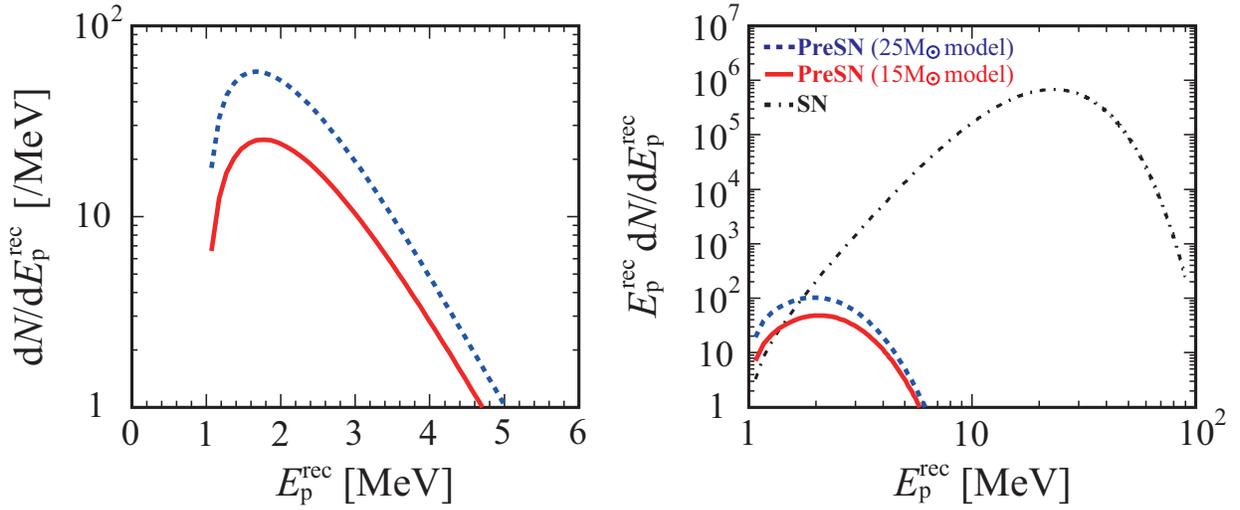}
\end{center}
\caption{
PreSN $\bar{\nu}_e$ event spectrum on Earth integrating over the last 48~hr preceding the collapse, with assuming a distance of 200~pc and the perfect 1~kt detector. 
Both panels are basically same with the linear-scale horizontal axis (left) and the log-scale horizontal axis (right). 
The vertical axis in the right panel is weighted by $E_{\rm p}^{\rm rec}$ to be $\sim dN/d\log E_{\rm p}^{\rm rec}$. 
The total number of preSN events in the detector is 44 and 95 for the 15~$M_{\sun}$ star and the 25~$M_{\sun}$ at 200~pc, respectively. 
In the right panel, for comparison, the weighted supernova $\bar{\nu}_e$ event spectrum integrating over 10~s are also shown. 
Expected number of SN events is about $8 \times 10^5$. The effect of neutrino oscillation is not considered.}
\label{fig3}
\end{figure}

In detectors, the reconstructed prompt~(positron in IBD) spectrum can be written as,
\begin{equation}\label{eq_convolution_integral}
 \frac{d^2N_M(t, E^{\rm rec}_{\rm p};d)}{dtdE^{\rm rec}_{\rm p}} = \epsilon_{\rm live} \epsilon_{\rm s}(E^{\rm rec}_{\rm p}) 
N_{\rm T}\int  dE^{\rm exp}_{\rm p} \phi_M(t,E_{\bar{\nu}_e};d) R(E^{\rm rec}_{\rm p},E^{\rm exp}_{\rm p})\sigma(E_{\bar{\nu}_e}),                      
\end{equation}
where $N_{\rm T}$ is the number of target protons in the analysis volume, 
$\sigma(E_{\bar{\nu}_e})$ is the IBD cross section~\citep{Strumia200342}, 
$\epsilon_{\rm live}$ is the mean livetime-to-runtime ratio, 
$\epsilon_{\rm s}(E^{\rm rec}_{\rm p})$ is the total detection efficiency. 
The details of these parameters are presented in Section \ref{sec:detector_and_background}.
$E^{\rm rec}_{\rm p}$ is the reconstructed energy and $E^{\rm exp}_{\rm p}=E_{\bar{\nu}_e}-0.78$\,MeV is the expected energy of the prompt event 
for an input $\bar{\nu}_e$ with an energy of $E_{\bar{\nu}_e}$. 
The integration in Equation (\ref{eq_convolution_integral}) is a convolution of the theoretical spectrum $\phi_M$ with the detector response. 
We model the detector response as a Gaussian with  $R(E^{\rm rec}_{\rm p},E^{\rm exp}_{\rm p})$ with energy resolution $\sigma_{\rm E}$:
\begin{equation}                                                                                                  
  R(E^{\rm rec}_{\rm p},E^{\rm exp}_{\rm p}) = \frac{1}{\sqrt{2\pi} \sigma_{\rm E}(E)}
\exp \Biggl\{ -\frac{ (E^{\rm rec}_{\rm p}-E^{\rm exp}_{\rm p})^2}{2 \sigma_{\rm E}^{2}(E)} \Biggr\}. 
\end{equation}
Assuming preSN emitted from the 200~pc star and the perfect 1~kt detector with $\epsilon_{\rm live}=1$, $\epsilon_{\rm s}(E^{\rm rec}_{\rm p})=1$, 
and $R(E^{\rm rec}_{\rm p},E^{\rm exp}_{\rm p})=\delta(E^{\rm rec}_{\rm p}-E^{\rm exp}_{\rm p})$, the event spectrum integrated over the last 48~hr preceding collapse is shown in the left panel of Figure~\ref{fig3}. 
The number of arrival $\bar{\nu}_e$ in the last 48~hr is $1.8 \times 10^{13}$ for the $15~M_\sun$ star and  $6.5 \times 10^{13}$ for the $25~M_\sun$ star, respectively. 
The total number of preSN events in the detector is 44 and 95, respectively. 

The right panel of Figure~\ref{fig3} shows the preSN spectrum integrating over the last 48~hr 
and SN spectrum integrating over 10~sec with the vertical axis of $E_{\rm p}^{\rm rec} dN/dE_{\rm p}^{\rm rec} \sim  dN/d\log E_{\rm p}^{\rm rec}$. 
Our assumption of supernovae at 200~pc is 50 times closer than the usually assumed 10~kpc. 
We note that the SN from 200~pc supernovae will create $\sim 10^6$ events in the detector. 
The current KamLAND electronics will not be able to record more than the basic hit information for the SN. 
The information from the preSN will not be lost in the case of the DAQ crash. 

Neutrino oscillation in the preSN emission region reduces the $\bar{\nu}_e$ flux. 
The flux of observable $\bar{\nu}_e$ can be expressed following \cite{Kneller2008} as, 
\begin{equation}
 \phi_{\bar{\nu}_e} =  p \phi^0_{\bar{\nu}_e} + (1 - p) \phi^0_{\bar{\nu}_x}, 
\end{equation}
where $\phi^0_{\bar{\nu}_e}$ and $\phi^0_{\bar{\nu}_x}$ are the original spectra of $\bar{\nu}_e$ and $\bar{\nu}_{\mu,\tau}$. 
With the assumption of $\phi^0_{\bar{\nu}_x} = 0.19\phi^0_{\bar{\nu}_e}$ based on $\phi^0_{\bar{\nu}_x} \propto \phi^0_{\bar{\nu}_e}$ and a $\bar{\nu}_x/\bar{\nu}_e$ ratio of 0.19 from \cite{Odrzywolek2004}, 
and an adiabatic approximation for $p$ with $\sin^2\theta_{12} = 3.08 \times 10^{-1}$ and $\sin^2\theta_{13} = 2.34 \times 10^{-2}$~($2.40\times 10^{-2}$) from \cite{Capozzi2014}, 
we have $\phi_{\bar{\nu}_e} = a \phi^0_{\bar{\nu}_e}$, where $a=0.74~(0.21)$ corresponding to the normal~(inverted) neutrino mass order. 
The corrected spectrum for the neutrino oscillation is then given by multiplying the coefficient $a$. 
With the integral over the last 48~hr preceding collapse, the event spectrum is shown with the oscillation effect and the full detector response in the middle and bottom panels of Figure~\ref{fig4} with KamLAND using the parameters describe in Sec.~\ref{sec:detector_and_background}.

\section{KamLAND detector and its background}\label{sec:detector_and_background}
KamLAND is located in the Kamioka Mine in Japan's Gifu prefecture ($36.42^{\circ}$N, $137.31^{\circ}$E). Mt.~Ikenoyama rises $\sim 1$\,km above the detector reducing backgrounds due to cosmic rays by five orders of magnitude. 
The KamLAND detector consists of approximately 1~kt of liquid scintillator, a mixture of 20\% psuedocume and 80\% dodecane. 
It is contained in a 13-m-diameter spherical balloon made of
a 135-$\mu$m-thick transparent nylon ethylene vinyl alcohol copolymer~(EVOH) composite film.  An array of photomultiplier tubes~(PMTs) is used to detect the scintillation light from events occurring within the balloon.  This array consists of 1,325 fast PMTs masked to 17-inch diameter to achieve the desired timing performance and
554 older 20-inch diameter PMTs reused from the Kamiokande experiment. The PMTs are mounted on the inner surface of an 18~m-diameter stainless steel sphere. Non-scintillating mineral oil fills the space between the balloon and the inner surface of the sphere. Its density is matched to the liquid scintillator to support the balloon and also acts as a passive shielding against external backgrounds from the sphere, PMTs, and surrounding rocks. This inner detector is further shielded by a 3.2\,kton water-Cherenkov veto detector.  
In 2011, a 3.08~m-diameters inner balloon containing 13~tons of Xe-loaded liquid scintillator (Xe-LS) was installed in the center of the main balloon as a part of the KamLAND Zero-Neutrino Double-Beta Decay~(KamLAND-Zen) experiment~\citep{kamlandzen2012}. 

The position and energy of an event within the balloon can be reconstructed 
using the timing and charge distribution obtained from the PMT array.  The reconstruction is calibrated by a number of radioactive sources: $^{60}$Co, $^{68}$Ge, $^{203}$Hg, $^{65}$Zn, $^{241}$Am$^{9}$Be, $^{137}$Cs, and $^{210}$Po$^{13}$C~\citep{Berger2009,Banks201588}.  From these calibrations and naturally occurring radioactive sources, the energy resolution ($\sigma_{\rm E}(E)$)is determined to be 6.4$\%/ \sqrt{E~{\rm (MeV)}}$ and the position reconstruction 12~cm$/ \sqrt{E~{\rm (MeV)}}$. 

Candidate $\bar{\nu}_e$ events corresponding to the prompt positron annihilation and delayed neutron capture of the IBD interaction are selected with a series of cuts on the energy, position, time and space coincidence of the two events. The two events must occur with $0.5<\Delta T$($\mu$s)$<$1000 and within $\Delta R<$2.0~m, where $\Delta T$ and $\Delta R$ are time and spatial differences. The reconstructed position of both events must be within a spherical fiducial volume $R_{\rm p}, R_{\rm d} <6$~m, which determines the fiducial number of target proton, $N_{\rm T}=5.98 \times 10^{31}$. The reconstructed energy of the prompt event is required to be in the energy range $E^{\rm rec}_{\rm p} ({\rm MeV}) \geq 0.9$. The delayed event has an energy characterized by the energy of the neutron capture gammas. Two energies are used:  $1.8< E^{\rm rec}_{\rm d} ({\rm MeV}) < 2.6$ corresponding to capture on H and $4.4 < E^{\rm rec}_{\rm d}({\rm MeV}) < 5.6$  corresponding to capture on $^{12}$C. An energy and position dependent likelihood variable is constructed to differentiate $\overline{\nu}_e$ from backgrounds due to accidental coincidences, which become more likely at lower energies and as events reconstruct closer to the balloon~\citep{kamland2011}. Finally, an additional position cut on the delayed event is applied to eliminate backgrounds due to the KamLAND-Zen inner balloon and 
support structure. The cut eliminates a central sphere and cylinder:   $R_{\rm d}<2.5$~m and $\sqrt{x^2_{\rm d}+y^2_{\rm d}}<2.5$~m for $z_{\rm d}> 0$~m where $(x_{\rm d}, y_{\rm d}, z_{\rm d})$ is the reconstructed position of the delayed capture event.

\begin{figure}
\begin{center}
\includegraphics[width=1.0\linewidth]{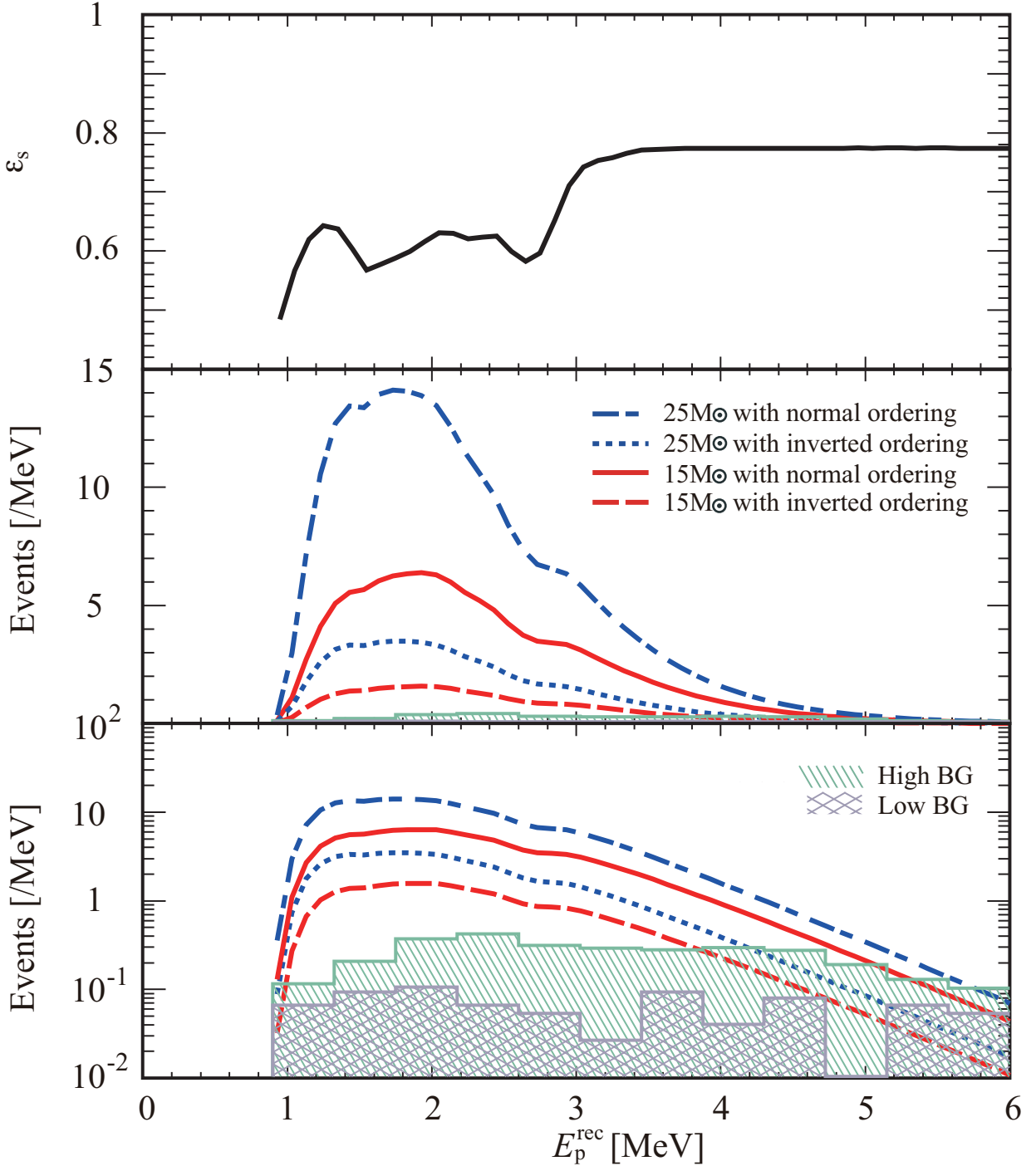}
\end{center}
\caption{
(Top) The energy dependent detection efficiency $\epsilon_{\rm s} (E^{\rm rec}_{\rm p})$ used in the analysis. 
(Middle) The integrated preSN energy spectrum and measured background in the last 48 hr before collapse. 
The backgrounds are dominated by reactor and geological $\overline{\nu}_e$. 
Background levels for the low-reactor phase~($B_{\rm low}$) and high-reactor phase~($B_{\rm high}$) are shown. 
(Bottom) Same as the middle panel but shown on a log scale instead, which clearly show the background spectra. 
Our analysis range is 0.9 -- 3.5\,MeV in the prompt energy $E_{\rm p}^{\rm rec}$.  }
\label{fig4}
\end{figure}

This series of cuts matches the standard KamLAND analysis~\citep{kamland2013}. 
The minimum prompt energy is chosen to guarantee 100\% detection efficiency for $\bar{\nu}_e$ with a energy of 1.8~MeV.   
The total efficiency of these cuts, $\epsilon_{\rm s} (E^{\rm rec}_{\rm p})$, is energy dependent due to the Likelihood selection as shown in Figure~\ref{fig4} (Top). 
The efficiency loss is dominated by the inner balloon cut. Without this cut, the efficiency is higher, $\sim$0.9 depending on the energy of the prompt event.  Further cuts are used to remove backgrounds due to high energy muon events. The effect of these cuts is to reduce the effective livetime to $\epsilon_{\rm live}=0.903$ even though KamLAND takes data continuously.

After these cuts are applied, the measured event spectrum in the preSN region ($0.9 \leq E^{\rm rec}_{\rm p} ({\rm MeV}) \leq 3.5$) 
is mainly from reactor $\overline{\nu}_e$ and geological $\overline{\nu}_e$ produced in the Earth's interior. 
These are the backgrounds to the preSN signal. 
Since the Great East Japan Earthquake, the reactors in Japan have been off.
This is the low-reactor phase, and the reduced backgrounds increase our preSN sensitivity. 
The measured background spectra are shown in the middle and bottom panels in Figure~\ref{fig4} for 
the low-reactor phase and the high-reactor phase. In this figure, the background spectra are normalized to the 48~hr window.

\section{Sensitivity}
\begin{figure}
\begin{center}
\includegraphics[width=1.0\linewidth]{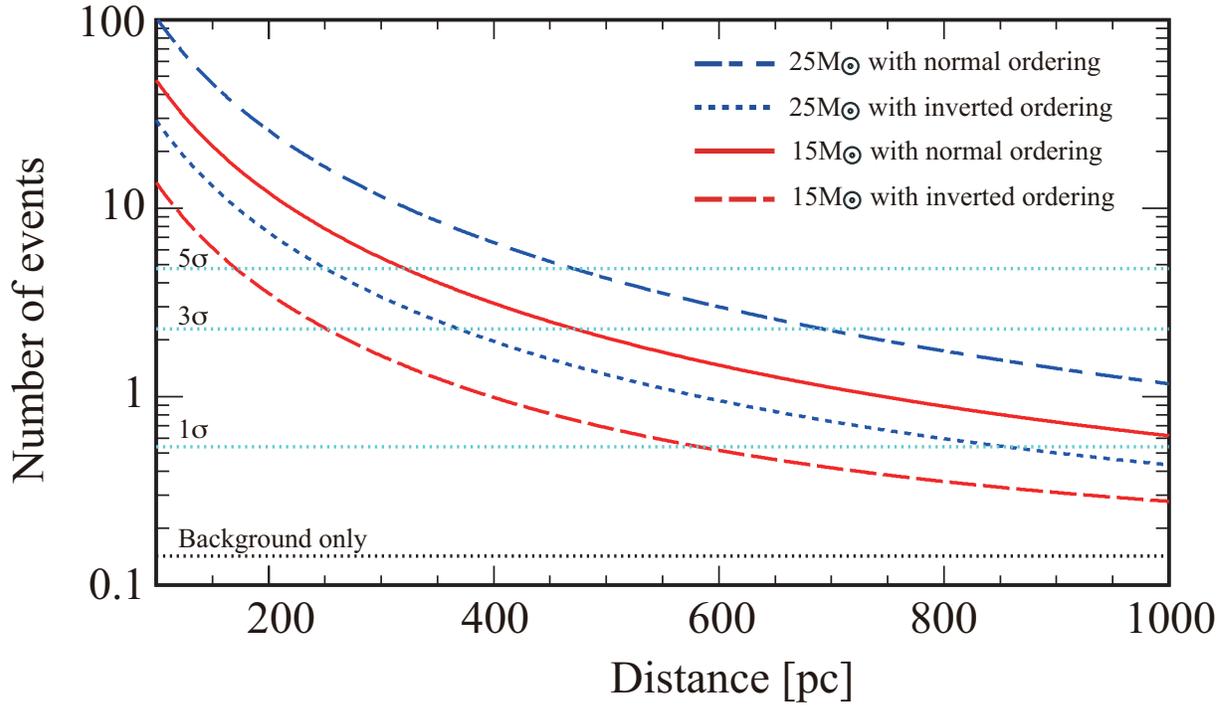}
\end{center}
\caption{Expected number of IBD events detected in KamLAND during the 48~hr before collapse including neutrino oscillation effects as a function of distance.  The flux for a 15\,$M_{\rm \sun}$ and  25\,$M_{\rm \sun}$ star is shown assuming the normal and inverted neutrino mass ordering. Horizontal dotted lines are the significance of the detection (see text).}
\label{fig5}
\end{figure}

The middle and bottom panels in Figure~\ref{fig4} show measured background and expected preSN spectra, 
integrated over the 48~hr window immediately before the collapse. 
Different integration lengths were studied and this length was chosen to maximize the signal-to-background ratio. 
To study the sensitivity, we use the analysis range of  $0.9 \leq E^{\rm rec}_{\rm p} ({\rm MeV}) \leq 3.5$ 
to maximize the preSN signal to background ratio while retaining~$>$90\% of the preSN signal. 
The background rate is $B_{\rm low} = 0.071$~event/day in the low-reactor phase. 
If the reactors return to normal operations, the background rate rises to $B_{\rm high} = 0.355$~events/day. 
The average efficiency is $\langle \epsilon_{\rm s} \rangle=0.64$ in this range. 
For the number of events $N_{M}^Z(d)=aN_M(d)$, where ${\rm Z}$ indicates the neutrino mass ordering~(${\rm Z}={\rm N/I}$ for the normal/inverted order), 
we integrate Eq.~(\ref{eq_convolution_integral}) from 0.9~MeV to 3.5~MeV and over the 48~hr before the collapse. 
They are now $N_{15M_{\sun}}^{\rm N}(200~{\rm pc})=12.0$, $N_{25M_{\sun}}^{\rm N}(200~{\rm pc})=25.7$ 
and $N_{15M_{\sun}}^{\rm I}(200~{\rm pc})=3.38$, $N_{25M_{\sun}}^{\rm I}(200~{\rm pc})=7.28$. 
The $N_{M}^{\rm Z}(d)$ are shown as a function of $d$ in Figure \ref{fig5}. 

Using $N_{M}^{\rm Z}(d)$ with background 2~days$ \times B_i$ ($i$=low or high), 
the corresponding detection significance, expressed in sigma, is estimated. 
The results for a detection significance of $1\sigma$, $3\sigma$, and $5\sigma$ assuming $B_{\rm low}$ are plotted in Figure \ref{fig5}. If a $3\sigma$ significance is required for preSN detection, KamLAND is sensitive to preSN from a 25~$M_{\sun}$ star at 690~pc assuming $B_{\rm low}$ and the normal neutrino mass ordering. In the worst case, KamLAND is still sensitive to a 15~$M_{\rm \sun}$ star at 250~pc. In this distance range, there are several red-supergiants which could lead to supernovae: Antares (150~pc), Betelgeuse (200~pc), Epsilon Pegasi (210~pc), Pi Puppis (250~pc), Sigma Canis Majoris (340~pc), NS Puppis (520~pc), CE Tauri (550~pc) and 3 Ceti (640~pc). 

\begin{figure}
\begin{center}
\includegraphics[width=1.0\linewidth]{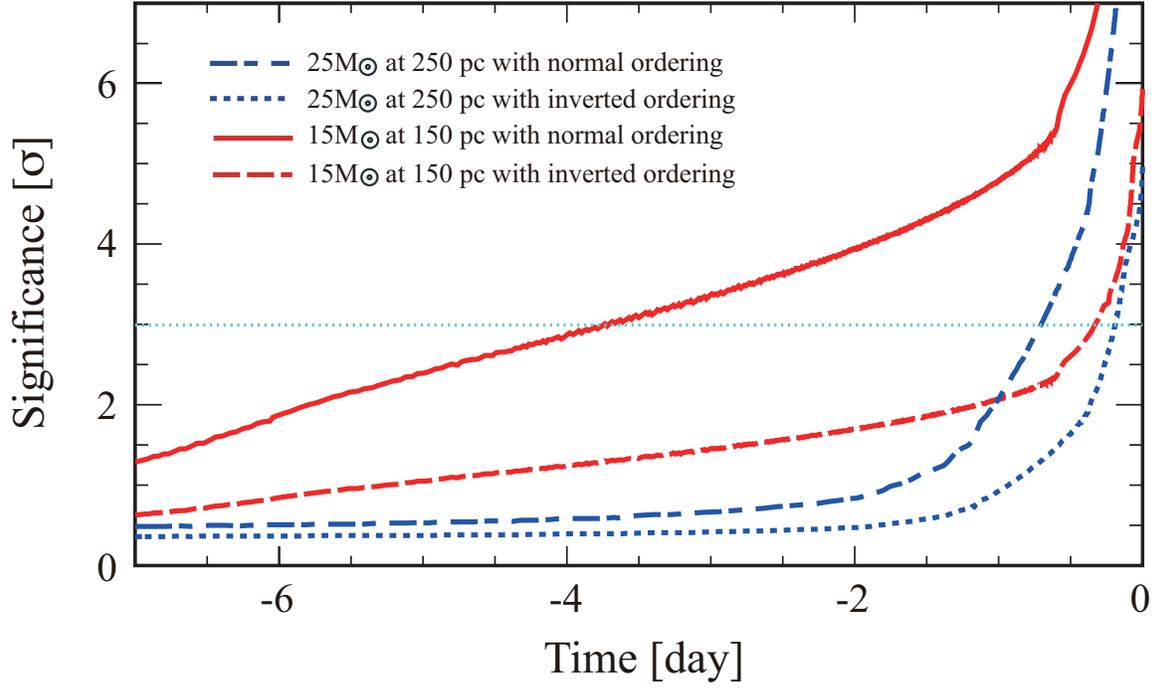}
\end{center}
\caption{Time evolution of significance in the low-reactor phase. 
If Betelgeuse has a mass of 15\,$M_{\rm \sun}$ at $d=150$\,pc, 
the 3$\sigma$ detection time (shown by a dotted horizontal line) is 89.6~(7.41)~hr with the normal~(inverted) ordering before collapse.  KamLAND can detect preSN 17.0 (4.54) hr before the collapse of Betelgeuse~(25~$M_{\rm \sun}$, $250$~pc) at the 3$\sigma$ level.}
\label{fig6}
\end{figure}

Betelgeuse has been studied extensively as a nearby pre-supernova star, see \cite{townes2009,Haubois2009,Ohnaka2009}, therefore we use it to determine KamLAND's sensitivity as a function of time before collapse. Betelgeuse's measured mass $M=17$--$25~M_{\sun}$ and distance $d=197 \pm 45$~pc are highly correlated~\citep{harper2009}.  We studied the two extreme cases: ($M$, $d$) = (15~$M_{\rm \sun}$, $150$~pc) and (25~$M_{\rm \sun}$, $250$~pc). The expected time evolution of significance with the 48~hr integration window is shown in Figure~\ref{fig6}, assuming the low-reactor background. 

If Betelgeuse has ($M$, $d$) =(15~$M_{\rm \sun}$, $150$~pc), KamLAND will easily detect its preSN. 
A 3$\sigma$ detection of preSN would be 89.6~(7.41) hr before collapse for the normal~(inverted) mass ordering. If Betelgeuse has (25~$M_{\rm \sun}$, $250$~pc), the increased distance reduces the preSN flux and the number of hours before collapse which KamLAND could detect preSN. If the reactors in Japan are restarted, the number of hours is also reduced because of the larger backgrounds. Table~\ref{tab:time} summarizes the results and shows for all of these cases that 
KamLAND can still detect preSN and has the ability to send a supernova alarm before collapse.

\section{Discussion}
\label{discussion}
We quantitatively evaluated KamLAND's sensitivity to preSN. Using Betelgeuse as a likely progenitor, we find that a $3\sigma$ detection of preSN 
at 2--90~hr before the collapse is possible. This a great improvement over the warning time provided by the current SuperNova Early Warning System (SNEWS) described in \cite{SNEWS}. 
The SNEWS alarm is not sufficient to bring gravitational wave and neutrino detectors back online from commissioning or calibration campaigns~\citep{Dooley:2014nga} 
or redirect telescope time for shock breakout observations~\citep{Adams:2013ana}. 
In comparison, the 2 hour to days provided by KamLAND 's detection of preSN would facilitate these measurements.

This is the motivation for the development of the preSN alert system. It provides two levels of alarms. A low-level alarm is produced using the semi-realtime preSN detection significance. This is calculated using a 48~hr integration window and the background level averaged over the past three months. A new window is opened every 15 min and has a latency of 25 min due to KamLAND's online data processing time. This system requires users to sign up to receive the current significance of detection, see the KamLAND web site\footnote{\tt http://www.awa.tohoku.ac.jp/kamland/SNmonitor/regist/index.html}.  A high-level alarm will report any 5$\sigma$ detections to 
the The Gamma-ray Coordinates Network~(GCN) and/or the Astronomer's Telegram~(ATel). This alarm is only sent after that the collaboration rejects other possibilities such as DAQ problems, mine activity, and radon contamination. Unfortunately, the IBD reaction does not provide directional information so a definitive localization requires 
SNEWS alarms, direction detection in Super-Kamiokande,  
coherent network analysis of gravitational waves~\citep{hayama2015} and/or electromagnetic observations. However, the limited number of 
known local progenitors limits the total number of targets that could create a preSN signal in KamLAND and therefore some localization is still possible.

In addition to the alarm, preSN could teach us about neutrino and supernova properties. The detection of preSN from Betelgeuse's supernova and an improvement in the measurement of its mass and distance would allow the determination of the normal neutrino mass ordering at 3.6$\sigma$
~(2.5$\sigma$) for the 25~$M_{\sun}$ (15~$M_{\sun}$) star. 
The detection of preSN could also distinguish between a supernova with an ONe core or an Fe core~\citep{kato2015}.

\section{Summary}
We evaluated the KamLAND's sensitivity to preSN, neutrino-antineutrino pairs from massive stars that have entered the late burning phases. 
Detection of preSN could provide an early warning for the imminent occurrence of a supernova 
and provide an opportunity to study stellar evolution models in the last stages prior to collapse.  This study quantitatively confirms that the $3\sigma$ detection of preSN is possible for stars with distances up to 690 pc under optimal conditions. The number of pre-supernova stars at this distance is limited but includes several promising candidates such as Betelgeuse. KamLAND presently provides the community with a preSN alarm based on the semi-realtime significance calculation and will send a report to GCN/ACTel for any 5$\sigma$ detection that have been verified by the KamLAND collaboration. 

There are several other detectors under construction or proposed: SNO+~\citep{sno+}, RENO-50, HANOHANO~\citep{hanohano}, JUNO~\citep{juno}, LENA~\citep{lena2012}, ASDC~\citep{asdc2014}, and Gd-doped Super-Kamiokande~\citep{GADZOOKS}. These detectors are expected to have similar or higher sensitivity to preSN and a coincident preSN search would significantly reduce false signals. 
A future combined alarm system could increase the detection range to a few kpc. This extended range would include several other pre-supernova stars including Eta-Carinae.  

The preSN are an exciting tool for both the study of stellar evolution and the supernova observation. The current models of preSN production are fairly simple.  KamLAND's sensitivity to both the flux and energy of the preSN could be used to extract more information on the late stages of evolution 
if more detailed predictions become available.

\acknowledgments
The authors thank A.~Odrzywolek for providing his data.  
KamLAND is supported by MEXT KAKENHI Grant Numbers 26104002, 26104007;  the World Premier International Research Center Initiative (WPI Initiative), MEXT, Japan; and under the U.S. Department of Energy (DOE) grants no. DE-FG03-00ER41138, DE-AC02-05CH11231, and DE-FG02-01ER41166, as well as other DOE and NSF grants to individual institutions, and Stichting Fundamenteel Onderzoek der Materie (FOM) in the Netherlands. The Kamioka Mining and Smelting Company has provided service for activities in the mine. We thank the support of NII for SINET4. 

\bibliographystyle{apj}
\bibliography{preSN}

\begin{thebibliography}{42}
\expandafter\ifx\csname natexlab\endcsname\relax\def\natexlab#1{#1}\fi

\bibitem[{Adams {et~al.}(2013)Adams, Kochanek, Beacom, Vagins, \&
  Stanek}]{Adams:2013ana}
Adams, S.~M., Kochanek, C., Beacom, J.~F., Vagins, M.~R., \& Stanek, K. 2013,
  \apj, 778, 164

\bibitem[{Agafonova {et~al.}(2015)}]{LVD}
Agafonova, N., {et~al.} 2015, \apj, 802, 47

\bibitem[{Alekseev {et~al.}(1987)Alekseev, Alekseeva, Volchenko, \&
  Krivosheina}]{baksan}
Alekseev, E.~N., Alekseeva, L.~N., Volchenko, V.~I., \& Krivosheina, I.~V.
  1987, J. Exp. Theor. Phys. Lett., 45, 589

\bibitem[{Alonso {et~al.}(2014)Alonso, Barros, Bergevin, Bernstein, Bignell,
  Blucher, Calaprice, \& Conrad}]{asdc2014}
Alonso, J.~R., Barros, N., Bergevin, M., {et~al.} 2014, arXiv:1409.5864

\bibitem[{Antonioli {et~al.}(2004)Antonioli, Fienberg, Fleurot, Fukuda,
  Fulgione, Habig, Heise, McDonald, Mills, Namba, Robinson, Scholberg,
  Schwendener, Sinnott, Stacey, Suzuki, Tafirout, Vigorito, Viren, Virtue, \&
  Zichichi}]{SNEWS}
Antonioli, P., Fienberg, R.~T., Fleurot, F., {et~al.} 2004, New Journal of
  Physics, 6, 114

\bibitem[{Arceo-D\'iaz {et~al.}(2015)Arceo-D\'iaz, Schr{\"o}der, Zuber, \&
  Jack}]{Arceo20151}
Arceo-D\'iaz, S., Schr{\"o}der, K.-P., Zuber, K., \& Jack, D. 2015,
  Astroparticle Physics, 70, 1

\bibitem[{Banks {et~al.}(2015)Banks, Freedman, Wallig, Ybarrolaza, Gando,
  Gando, Ikeda, Inoue, Kishimoto, Koga, Mitsui, Nakamura, Shimizu, Shirai,
  Suzuki, Takemoto, Tamae, Ueshima, Watanabe, Xu, Yoshida, Yoshida, Kozlov,
  Grant, Keefer, Piepke, Bloxham, Fujikawa, Han, Ichimura, Murayama,
  O׳Donnell, Steiner, Winslow, Dwyer, McKeown, Zhang, Berger, Lane, Maricic,
  Miletic, Batygov, Learned, Matsuno, Sakai, Horton-Smith, Downum, Gratta,
  Efremenko, Perevozchikov, Karwowski, Markoff, Tornow, Heeger, Detwiler,
  Enomoto, \& Decowski}]{Banks201588}
Banks, T., Freedman, S., Wallig, J., {et~al.} 2015, Nucl.~Instrum.~Meth.~A,
  769, 88

\bibitem[{Beacom \& Vagins(2004)}]{GADZOOKS}
Beacom, J.~F., \& Vagins, M.~R. 2004, Phys. Rev. Lett., 93, 171101

\bibitem[{Berger {et~al.}(2009)Berger, Busenitz, Classen, Decowski, Dwyer,
  Elor, Frank, Freedman, Fujikawa, Galloway, Gray, Heeger, Hsu, Ichimura,
  Kadel, Keefer, Lendvai, McKee, O'Donnell, Piepke, Steiner, Syversrud, Wallig,
  Winslow, Ebihara, Enomoto, Furuno, Gando, Ikeda, Inoue, Kibe, Kishimoto,
  Koga, Minekawa, Mitsui, Nakajima, Nakajima, Nakamura, Owada, Shimizu,
  Shimizu, Shirai, Suekane, Suzuki, Tamae, Yoshida, Kozlov, Murayama, Grant,
  Leonard, Luk, Jillings, Mauger, McKeown, Zhang, Lane, Maricic, Miletic,
  Batygov, Learned, Matsuno, Pakvasa, Foster, Horton-Smith, Tang, Dazeley,
  Downum, Gratta, Tolich, Bugg, Efremenko, Kamyshkov, Perevozchikov, Karwowski,
  Markoff, Tornow, Piquemal, \& Ricol}]{Berger2009}
Berger, B.~E., Busenitz, J., Classen, T., {et~al.} 2009, JINST, 4, P04017

\bibitem[{Bionta {et~al.}(1987)Bionta, Blewitt, Bratton, Casper, Ciocio, Claus,
  Cortez, Crouch, Dye, Errede, Foster, Gajewski, Ganezer, Goldhaber, Haines,
  Jones, Kielczewska, Kropp, Learned, LoSecco, Matthews, Miller, Mudan, Park,
  Price, Reines, Schultz, Seidel, Shumard, Sinclair, Sobel, Stone, Sulak,
  Svoboda, Thornton, van~der Velde, \& Wuest}]{IMB}
Bionta, R.~M., Blewitt, G., Bratton, C.~B., {et~al.} 1987, Phys. Rev. Lett.,
  58, 1494

\bibitem[{Cadonati {et~al.}(2002)Cadonati, Calaprice, \&
  Chen}]{Cadonati2002361}
Cadonati, L., Calaprice, F., \& Chen, M. 2002, Astroparticle Physics, 16, 361

\bibitem[{Capozzi {et~al.}(2014)Capozzi, Fogli, Lisi, Marrone, Montanino, \&
  Palazzo}]{Capozzi2014}
Capozzi, F., Fogli, G.\, L., Lisi, E., {et~al.} 2014, Phys. Rev. D, 89, 093018

\bibitem[{Chen(2008)}]{sno+}
Chen, M.~C. 2008, arXiv:0810.3694

\bibitem[{Dolan {et~al.}(2014)Dolan, Mathews, Lam, Lan, J, Herczeg, \&
  Dearborn}]{Dolan2014}
Dolan, M.~M., Mathews, G.~J., Lam, D.~D., {et~al.} 2014, arXiv:1406.3143

\bibitem[{Dooley(2015)}]{Dooley:2014nga}
Dooley, K.~L. 2015, J. Phys. Conf. Ser., 610, 012015

\bibitem[{Gando {et~al.}(2011)Gando, Gando, Ichimura, Ikeda, Inoue, Kibe,
  Kishimoto, Koga, Minek~awa, Mitsui, Morikawa, Nagai, Nakajima, Nakamura,
  Narita, Shimizu, Shimizu, Shirai, Suekane, Suzuki, Takahashi, Takahashi,
  Takemoto, Tamae, Watanabe, Xu, Yabumoto, Yoshida, Yoshida, Enomoto, Kozlov,
  Murayama, Grant, Keefer, Piepke, Banks, Bloxham, Detwiler, Freedman,
  Fujikawa, Han, Kadel, O'Donnell, Steiner, Dwyer, McKeown, Zhang, Berger,
  Lane, Marici~c, Miletic, Batygov, Learned, Matsuno, Sakai, Horton-Smith,
  Downum, G~ratta, Efremenko, Perevozchikov, Karwowski, Markoff, Tornow,
  Heeger, \& Decowsk~i}]{kamland2011}
Gando, A., Gando, Y., Ichimura, K., {et~al.} 2011, Phys. Rev. D, 83, 052002

\bibitem[{Gando {et~al.}(2012)Gando, Gando, Hanakago, Ikeda, Inoue, Kato, Koga,
  Matsuda, Mitsui, Nakada, Nakamura, Obata, Oki, Ono, Shimizu, Shirai, Suzuki,
  Takemoto, Tamae, Ueshima, Watanabe, Xu, Yamada, Yoshida, Kozlov, Yoshida,
  Banks, Detwiler, Freedman, Fujikawa, Han, O'Donnell, Berger, Efremenko,
  Karwowski, Markoff, Tornow, Enomoto, \& Decowski}]{kamlandzen2012}
Gando, A., Gando, Y., Hanakago, H., {et~al.} 2012, Phys. Rev. C, 85, 045504

\bibitem[{Gando {et~al.}(2013)Gando, Gando, Hanakago, Ikeda, Inoue,
  Ishidoshiro, Ishikawa, Koga, Matsuda, Matsuda, Mitsui, Motoki, Nakamura,
  Obata, Oki, Oki, Otani, Shimizu, Shirai, Suzuki, Takemoto, Tamae, Ueshima,
  Watanabe, Xu, Yamada, Yamauchi, Yoshida, Kozlov, Yoshida, Piepke, Banks,
  Fujikawa, Han, O'Donnell, Berger, Learned, Matsuno, Sakai, Efremenko,
  Karwowski, Markoff, Tornow, Detwiler, Enomoto, \& Decowski}]{kamland2013}
---. 2013, Phys. Rev. D, 88, 033001

\bibitem[{Harper {et~al.}(2008)Harper, Brown, \& Guinan}]{harper2009}
Harper, G.~M., Brown, A., \& Guinan, E.~F. 2008, \apj, 135, 1430

\bibitem[{Haubois {et~al.}(2009)Haubois, Perrin, Lacour, Verhoelst, Meimon,
  Mugnier, Thi\'{e}baut, Berger, Ridgway, Monnier, Millan-Gabet, \&
  Traub}]{Haubois2009}
Haubois, X., Perrin, G., Lacour, S., {et~al.} 2009, Astron. Astrophys., 508,
  923

\bibitem[{Hayama {et~al.}(2015)Hayama, Kuroda, Kotake, \&
  Takiwaki}]{hayama2015}
Hayama, K., Kuroda, T., Kotake, K., \& Takiwaki, T. 2015, arXiv:1501.00966

\bibitem[{Heger {et~al.}(2009)Heger, Friedland, Giannotti, \&
  Cirigliano}]{Heger2009}
Heger, A., Friedland, A., Giannotti, M., \& Cirigliano, V. 2009, \apj, 696, 608

\bibitem[{Hirata {et~al.}(1987)Hirata, Kajita, Koshiba, Nakahata, Oyama, Sato,
  Suzuki, Takita, Totsuka, Kifune, Suda, Takahashi, Tanimori, Miyano, Yamada,
  Beier, Feldscher, Kim, Mann, Newcomer, Van, Zhang, \& Cortez}]{hirata1987}
Hirata, K., Kajita, T., Koshiba, M., {et~al.} 1987, Phys. Rev. Lett., 58, 1490

\bibitem[{Hirata {et~al.}(1988)Hirata, Kajita, Koshiba, Nakahata, Oyama, Sato,
  Suzuki, Takita, Totsuka, Kifune, Suda, Takahashi, Tanimori, Miyano, Yamada,
  Beier, Feldscher, Frati, Kim, Mann, Newcomer, Van~Berg, Zhang, \&
  Cortez}]{hirata1988}
Hirata, K.~S., Kajita, T., Koshiba, M., {et~al.} 1988, Phys. Rev. D, 38, 448

\bibitem[{Itoh {et~al.}(1996)Itoh, Hayashi, Nishikawa, \& Kohyama}]{Itoh1996}
Itoh, N., Hayashi, H., Nishikawa, A., \& Kohyama, Y. 1996, \apjs, 1024, 411

\bibitem[{Kato {et~al.}(2015)Kato, Azari, Yamada, Takahashi, Umeda, Yoshida, \&
  Ishidoshiro}]{kato2015}
Kato, C., Azari, M.~D., Yamada, S., {et~al.} 2015, \apj, 808, 168

\bibitem[{Kim(2014)}]{Kim:2014rfa}
Kim, S.-B. 2014, arXiv:1412.2199

\bibitem[{Kneller {et~al.}(2008)Kneller, McLaughlin, \& Brockman}]{Kneller2008}
Kneller, J.~P., McLaughlin, G.~C., \& Brockman, J. 2008, \prd, 77, 045023

\bibitem[{Learned {et~al.}(2008)Learned, Dye, \& Pakvasa}]{hanohano}
Learned, J.~G., Dye, S.~T., \& Pakvasa, S. 2008, arXiv:0810.4975

\bibitem[{Li(2014)}]{juno}
Li, Y.-F. 2014, arXiv:1402.6143

\bibitem[{Nakazato {et~al.}(2013)Nakazato, Sumiyoshi, Suzuki, Totani, Umeda, \&
  Yamada}]{nakazato}
Nakazato, K., Sumiyoshi, K., Suzuki, H., {et~al.} 2013, \apjs, 205, 2

\bibitem[{Novoseltseva {et~al.}(2011)Novoseltseva, Boliev, Vereshkov,
  Volchenko, Volchenko, Dzaparova, Kochkarov, Kostyuk, Novoseltsev, Petkov,
  Striganov, \& Yanin}]{baksan2011}
Novoseltseva, R., Boliev, M., Vereshkov, G., {et~al.} 2011, Bull. Russ. Acad.
  Sci. Phys., 75, 419

\bibitem[{Odrzywolek \& Heger(2010)}]{Odrzywolek2010}
Odrzywolek, A., \& Heger, A. 2010, ACTA Physica Polonica B, 41, 1611

\bibitem[{Odrzywolek {et~al.}(2004)Odrzywolek, Misiaszek, \&
  Kutschera}]{Odrzywolek2004}
Odrzywolek, A., Misiaszek, M., \& Kutschera, M. 2004, Astropart.~Phys., 21, 303

\bibitem[{Ohnaka {et~al.}(2009)Ohnaka, Hofmann, Benisty, Chelli, Driebe,
  Millour, Petrov, \~Schertl, Stee, Vakili, \& Weigelt}]{Ohnaka2009}
Ohnaka, K., Hofmann, K.-H., Benisty, M., {et~al.} 2009, Astron. Astrophys.,
  503, 183

\bibitem[{Renshaw {et~al.}(2014)Renshaw, Abe, Hayato, Iyogi, Kameda, Kishimoto,
  Miura, Moriyama, Nakahata, Nakano, Nakayama, Sekiya, Shiozawa, Suzuki,
  Takeda, Takenaga, Tomura, Ueno, Yokozawa, Wendell, Irvine, Kajita, Kaneyuki,
  Lee, Nishimura, Okumura, McLachlan, Labarga, Berkman, Tanaka, Tobayama,
  Kearns, Raaf, Stone, Sulak, Goldhabar, Bays, Carminati, Kropp, Mine, Smy,
  Sobel, Ganezer, Hill, Keig, Hong, Kim, Lim, Akiri, Himmel, Scholberg, Walter,
  Wongjirad, Ishizuka, Tasaka, Jang, Learned, Matsuno, Smith, Hasegawa, Ishida,
  Ishii, Kobayashi, Nakadaira, Nakamura, Oyama, Sakashita, Sekiguchi,
  Tsukamoto, Suzuki, Takeuchi, Bronner, Hirota, Huang, Ieki, Ikeda, Kikawa,
  Minamino, Nakaya, Suzuki, Takahashi, Fukuda, Choi, Itow, Mitsuka, Mijakowski,
  Hignight, Imber, Jung, Yanagisawa, Ishino, Kibayashi, Koshio, Mori, Sakuda,
  Yano, Kuno, Tacik, Kim, Okazawa, Choi, Nishijima, Koshiba, Totsuka, Yokoyama,
  Martens, Marti, Vagins, Martin, de~Perio, Konaka, Wilking, Chen, Zhang, \&
  Wilkes}]{Renshaw2014}
Renshaw, A., Abe, K., Hayato, Y., {et~al.} 2014, Phys. Rev. Lett., 112, 091805

\bibitem[{Scholberg(2012)}]{kate2012}
Scholberg, K. 2012, Ann.~Rev.~Nucl.~Part.~Sci, 62, 81

\bibitem[{Strumia \& Vissani(2003)}]{Strumia200342}
Strumia, A., \& Vissani, F. 2003, Physics Letters B, 564, 42

\bibitem[{Townes {et~al.}(2009)Townes, Wishnow, Hale, \& Walp}]{townes2009}
Townes, C.~H., Wishnow, E.~H., Hale, D. D.~S., \& Walp, B. 2009, \apjl, 697,
  L127

\bibitem[{Vissani(2015)}]{Vissani2015}
Vissani, F. 2015, J. Phys. G: Nucl. Part. Phys, 42, 013001

\bibitem[{Woosley \& Heger(2015)}]{Woosley2015}
Woosley, S.~E., \& Heger, A. 2015, arXiv:1505.06712

\bibitem[{Wurm {et~al.}(2012)Wurm, Beacom, Bezrukov, Bick, Blümer, Choubey,
  Ciemniak, D’Angelo, Dasgupta, Derbin, Dighe, Domogatsky, Dye, Eliseev,
  Enqvist, Erykalov, von Feilitzsch, Fiorentini, Fischer, Göger-Neff,
  Grabmayr, Hagner, Hellgartner, Hissa, Horiuchi, Janka, Jaupart, Jochum,
  Kalliokoski, Kayunov, Kuusiniemi, Lachenmaier, Lazanu, Learned, Lewke,
  Lombardi, Lorenz, Lubsandorzhiev, Ludhova, Loo, Maalampi, Mantovani,
  Marafini, Maricic, Undagoitia, McDonough, Miramonti, Mirizzi, Meindl, Mena,
  Möllenberg, Muratova, Nahnhauer, Nesterenko, Novikov, Nuijten, Oberauer,
  Pakvasa, Palomares-Ruiz, Pallavicini, Pascoli, Patzak, Peltoniemi, Potzel,
  Räihä, Raffelt, Ranucci, Razzaque, Rummukainen, Sarkamo, Sinev, Spiering,
  Stahl, Thorne, Tippmann, Tonazzo, Trzaska, Vergados, Wiebusch, \&
  Winter}]{lena2012}
Wurm, M., Beacom, J.~F., Bezrukov, L.~B., {et~al.} 2012, Astropart.~Phys., 35,
  685

\end{thebibliography}

\clearpage

\clearpage

\begin{deluxetable}{ccccc}
\tablecaption{Expected time before the Betelgeuse supernova to reach 3$\sigma$ confidence based on PreSN under various parameter assumptions. \label{tab:time}} 
\tablewidth{0pt}
\tablehead{
\colhead{Mass~[$M_\sun$]} & \colhead{Distance~[pc]} & \colhead{Mass ordering} & \colhead{Reactor status} & \colhead{Time before collapse [hr]}  }
\startdata
   15 & 150 & Normal & low & 89.6   \\
   15 & 150 & Inverted & low & 7.41  \\
   25 & 250 & Normal & low & 17.0   \\   
   25 & 250 & Inverted & low & 4.54 \\
   15 & 150 & Normal & high & 46.0   \\
   15 & 150 & Inverted & high & 3.17  \\
   25 & 250 & Normal & high & 11.1  \\
   25 & 250 & Inverted & high & 1.93 \\\enddata
\\
\vspace{0.5 cm}
\end{deluxetable}

\end{document}